\documentclass[twocolumn]{IEEEtran}

\usepackage{graphicx}
\graphicspath{{figs/}}
\usepackage{scrextend} 
\usepackage{gensymb}
\addtokomafont{labelinglabel}{\sffamily}
\usepackage{multirow}
\usepackage[normalem]{ulem} 
\usepackage{todonotes}
\usepackage{amsmath}
\usepackage{xfrac}
\usepackage{url}
\usepackage{color} 
\usepackage{siunitx}

\usepackage[export]{adjustbox}
\ifCLASSOPTIONcompsoc
  \usepackage[caption=false, font=normalsize, labelfont=sf, textfont=sf]{subfig}
\else
  \usepackage[caption=false, font=footnotesize]{subfig}
\fi

\newcommand{\fref}[1]{Fig.\ #1}
\newcommand{\secref}[1]{Section #1}

\newcommand{\nm}{\nano\meter}
\newcommand{\meV}{\micro\eV}

\newcommand{\tts}[1]{\textsuperscript{#1}}  
\newcommand{\bpar}[1]{(\textbf{#1})}        

\DeclareMathOperator\erf{erf}

\begin{document}

\title{SiQAD: A Design and Simulation Tool for Atomic Silicon Quantum Dot Circuits}

\author{
  \IEEEauthorblockN{
    Samuel Ng,
    Jacob Retallick,
    Hsi Nien Chiu,
    Robert Lupoiu,
    Mohammad Rashidi,
    Wyatt Vine,
    Thomas Dienel,
    Lucian Livadaru,
    Robert A. Wolkow,
    Konrad Walus
  }
}

\IEEEtitleabstractindextext{%
  \begin{abstract}
    This paper introduces SiQAD, a computer-aided design tool enabling the rapid design and simulation of atomic silicon dangling bond quantum dot patterns capable of computational logic. Several simulation tools are included, each able to inform the designer on various aspects of their designs: a ground-state electron configuration finder, a non-equilibrium electron dynamics simulator, and an electric potential landscape solver with clocking electrode support. Simulations have been compared against past experimental results to inform the electron population estimation and dynamic behavior. New logic gates suitable for this platform have been designed and simulated, and a clocked wire has been demonstrated. This work paves the way for the exploration of the vast and fertile design space of atomic silicon dangling bond quantum dot circuits.
  \end{abstract}
}

\maketitle
\IEEEdisplaynontitleabstractindextext

\section{Introduction}

\newcommand{\hsi}{H-Si(100)-2$\times$1\ }

Recent advancements have made possible the precise creation and erasure of silicon dangling bonds (Si-DBs) on the hydrogen passivated Si(100) 2$\times$1 surface \cite{Huff2017_white-out,Pavlicek2017}. These Si-DBs have localized eigenstates in the band gap of the bulk and are interpreted as zero dimensional quantum dots with at most two occupying electrons \cite{Haider2009, Taucer2014}. The preferred charge configurations of specific patterns of Si-DBs enable the creation of logic gates \cite{Wolkow2014_logic, Huff2017_OR}, introducing a fundamentally new computational platform technology and unique design space at the atomic scale. Exploration of this prospective field requires new tools for simulating the behavior of Si-DB patterns.

This paper presents SiQAD (\textbf{Si}licon \textbf{Q}uantum \textbf{A}tomic \textbf{D}esigner), a CAD tool that provides an environment for the design and simulation of Si-DB circuits. \secref{\ref{sec:background}} provides the technical background on Si-DB quantum dot circuits. \secref{\ref{sec:siqad_overview}} introduces SiQAD's capability for circuit design and simulation, and describes the software architecture of the CAD tool. \secref{\ref{sec:simulation_engines}} provides detailed explanations and simulation results of the three included simulation tools. \secref{\ref{sec:new_gates}} presents new Si-DB logic gates designed and simulated in SiQAD.

\section{Background} \label{sec:background}

\begin{figure}
  \centering
  \subfloat[Surface of \hsi with Si-DB.]{%
    \includegraphics[width=.6\linewidth]{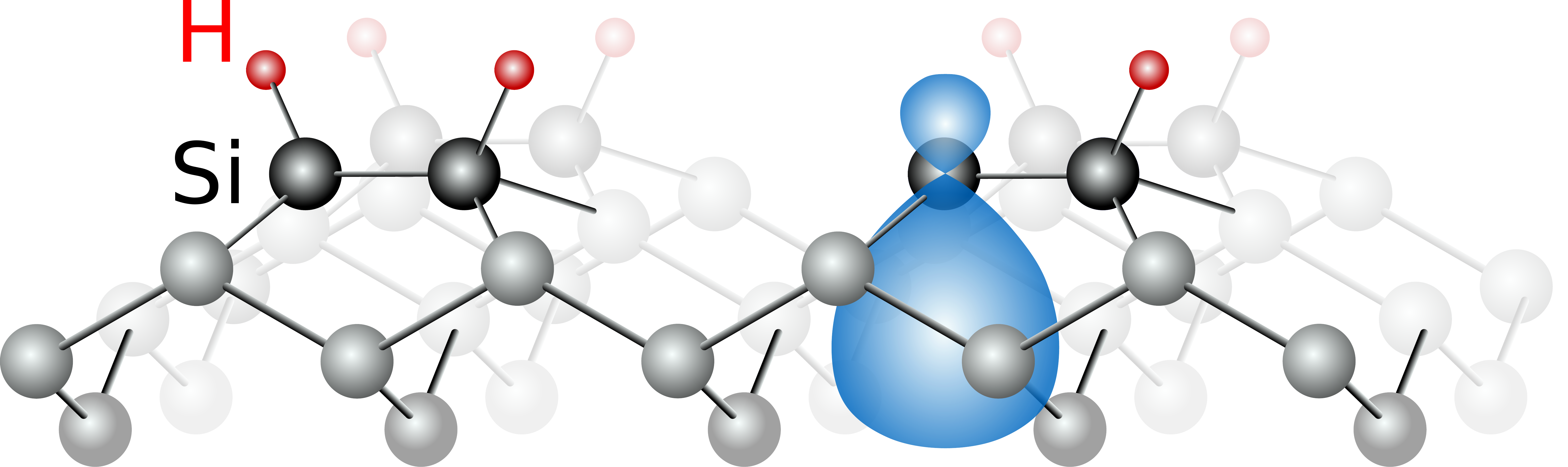}%
    \label{fig:sidb_orbital}}\quad
  \subfloat[Si-DB layout.]{%
    \includegraphics[width=.35\linewidth]{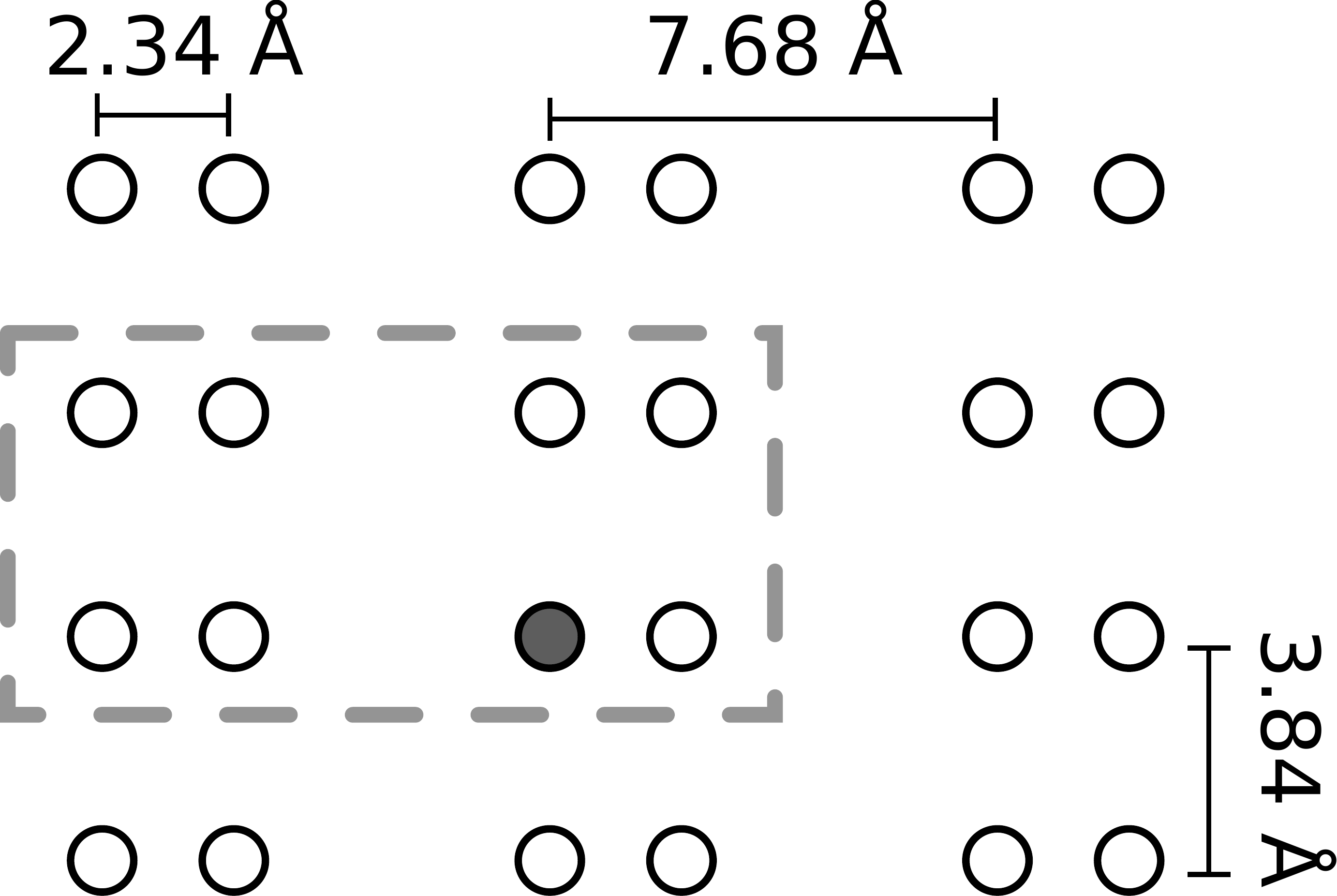}%
    \label{fig:sidb_layout}}\\[2ex]
  \subfloat[Schematic of Si-DB band structure.]{\includegraphics[width=.8\linewidth]{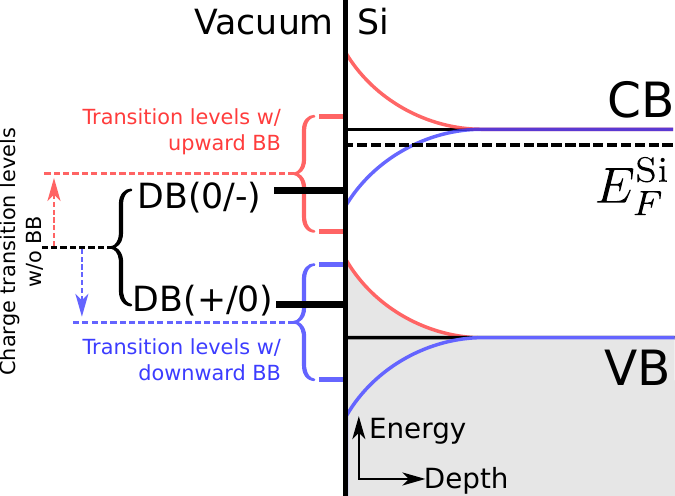} \label{fig:db_band_structure}}
  \caption{%
    \bpar{a} Desorbed hydrogen leaves a valence orbital in \hsi with mid-gap transition states. %
    \bpar{b} An Si-DB layout where the highlighted area shows a schematic top view of the region in (a). %
    \bpar{c} Schematic of the band structure of an Si-DB illustrating the mid-gap DB positive to neutral (+/0) and neutral to negative (0/-) transition states. The red, blue and black states represent upward band bending, downward band bending and no band bending respectively. The charge state of the Si-DB depends on the energetic position of its charge transition levels with respect to the bulk Fermi energy $E_{F}^{Si}$. %
  }
  \label{fig:Si-DB-schem}
\end{figure}

Si-DBs are fabricated by hydrogen desorption using the probe of a scanning tunneling microscope (STM), where H-Si bonds are broken by applying a voltage pulse at a hydrogen site \cite{Lyding1994, Shen1995, Foley1998}. The band structure of Si-DBs is illustrated in \fref{\ref{fig:Si-DB-schem}}. The unoccupied positive state (DB+), singly occupied neutral state (DB0) and doubly occupied negative state \mbox{(DB-)} exist within the band gap, allowing for discrete electron population of the surface DBs \cite{Haider2009, Taucer2014, Labidi2015, Livadaru2010}. Subsurface dopants, in addition to Coulombic interactions between Si-DBs and external fields such as the contact potential of the tip of a nearby STM or atomic force microscope (AFM), result in band bending \cite{Labidi2015, Rashidi2016_dope}. This band bending shifts the occupation levels with respect to the Fermi level, changing the preferred occupation of the Si-DBs as illustrated in \fref{\ref{fig:db_band_structure}}.

\begin{figure}
  \centering
  \includegraphics[width=\linewidth]{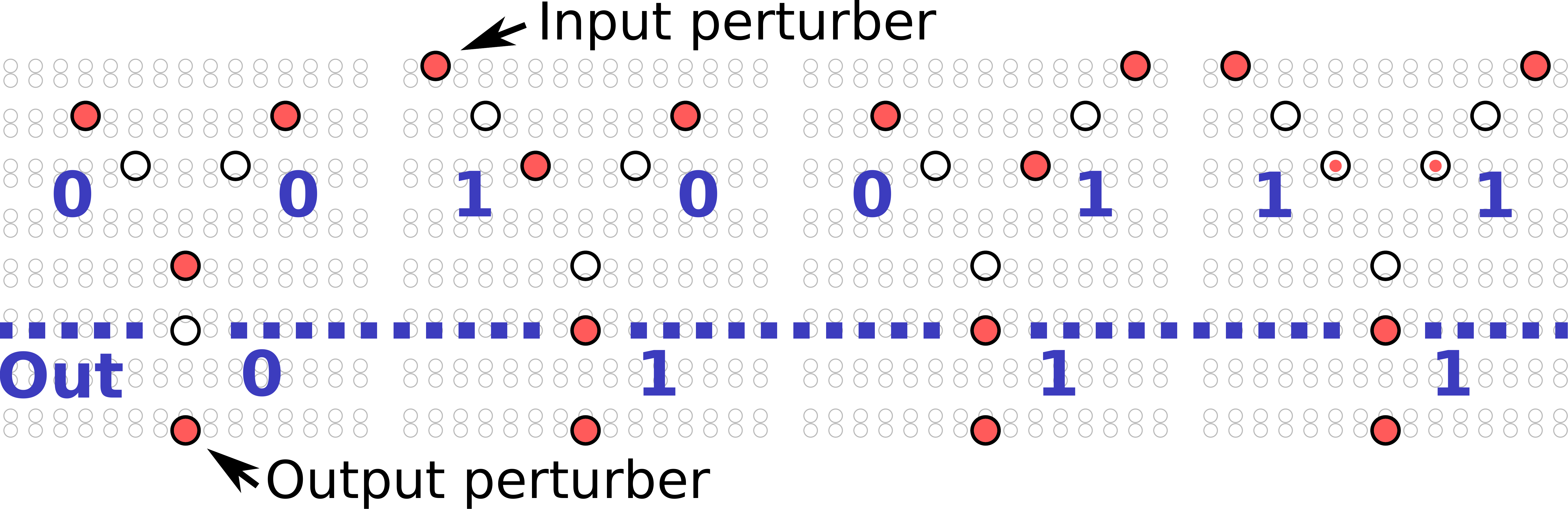}
  \caption{Simulated annealing results of an Si-DB OR gate made of 3 pairs of binary dots studied in \cite{Huff2017_OR} for all input scenarios. The logical output of the gate is determined by the charge state of the Si-DB labeled `Out': 0 and 1 for DB0 and DB- respectively. The presence of \textit{perturbers} at the top of the device change the input state. The output perturber acts to prevent a charge from occupying the output unless sufficiently repelled by charges in the upper half of the device. Simulated annealing was performed with $\mu=\SI{0.25}{\eV}$ (see~\eqref{eq:sa-pop}). Hollowed Si-DBs are singly occupied while the inclusion of a red fill indicates the average double-occupancy of each Si-DB over the last 1000 charge states in the annealing.}
  \label{sfig:simanneal_or}
\end{figure}

The potential of Si-DBs as building blocks for computational logic circuits has previously been demonstrated \cite{Huff2017_OR}, where a binary OR gate that consists of 3 central pairs of Si-DBs was proposed. \fref{\ref{sfig:simanneal_or}} shows a reproduction of the gate using SiQAD with simulation results that will be explained in \secref{\ref{sec:simulation_engines}}. The inputs are defined by the presence of \textit{perturbers}, peripheral Si-DBs tending to be in the DB- state, and the output is given by the charge state of a particular Si-DB. The charge states were measured via AFM in \cite{Huff2017_OR} exhibiting a tendency to relax to the system's ground state.

\begin{figure}
  \centering
  \hspace{0.05\linewidth}
  \subfloat{%
    \includegraphics[angle=90,width=.08\linewidth,valign=t]{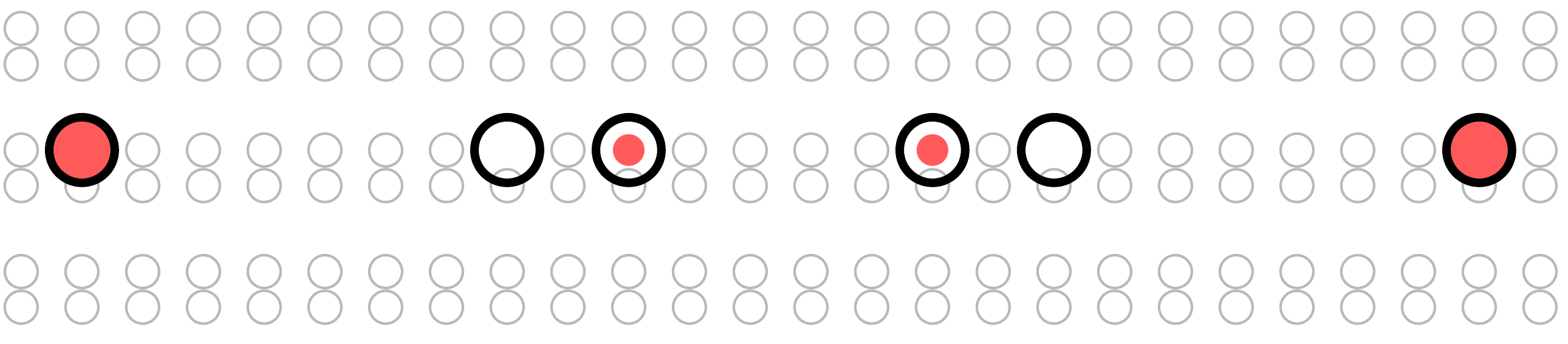}
    \label{sfig:simanneal_1-2-2-1}}\hfill
  \subfloat{%
    \includegraphics[width=.8\linewidth,valign=t]{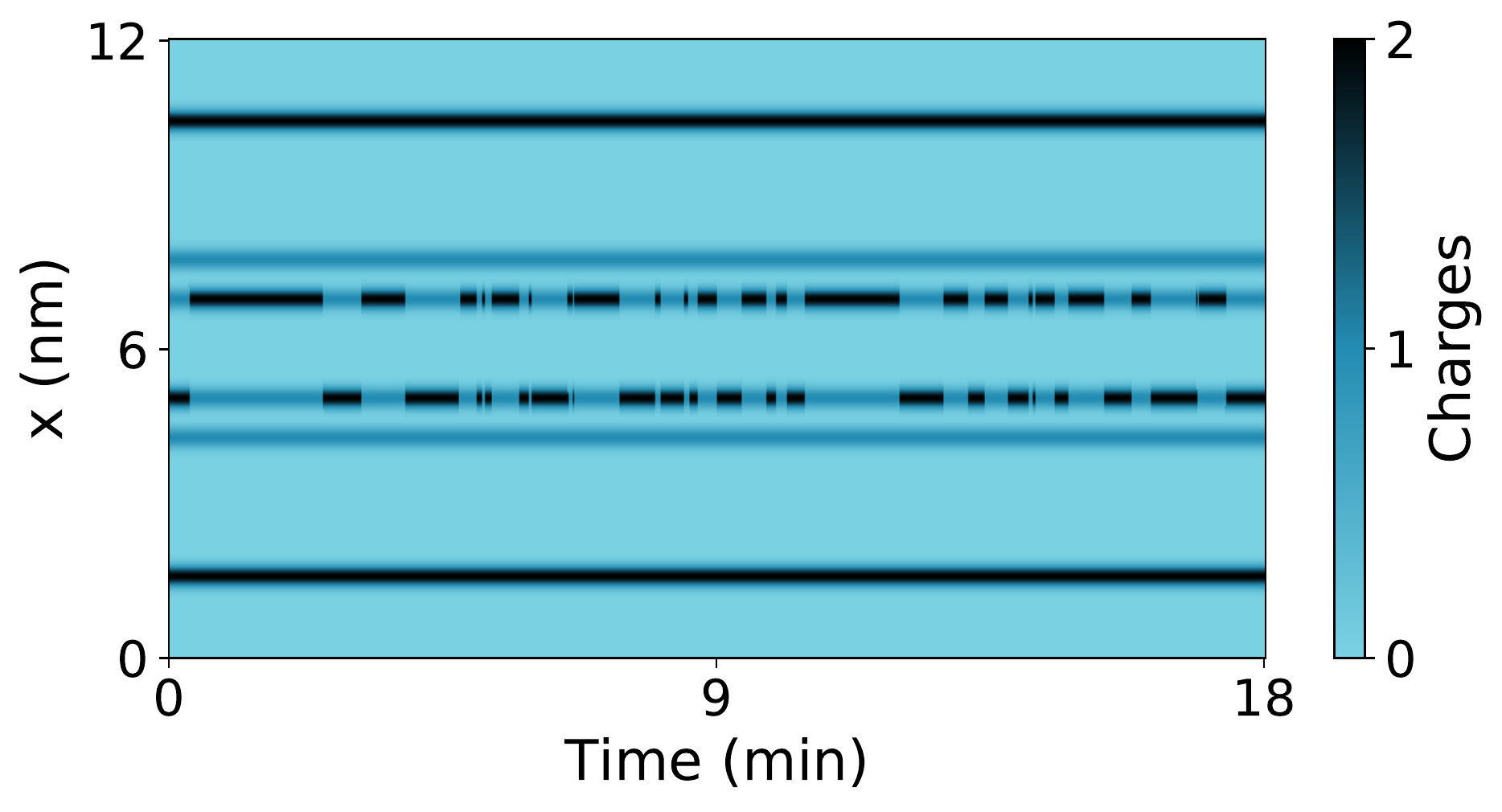}
    \label{sfig:hoppingdyn_1-2-2-1}}\hfill

  \caption{Six Si-DB symmetric structure studied in \cite{Rashidi2017_1-2-2-1}. %
    \bpar{left} Simulated average occupation with SimAnneal ($\mu = \SI{0.08}{\eV}$). %
    \bpar{right} Dynamic simulation of the structure with HoppingDynamics with 800 line scans over 18 minutes showing similar results to \cite{Rashidi2017_1-2-2-1}. %
    }
  \label{fig:qsi_1-2-2-1}
\end{figure}

Additionally, a six Si-DB symmetric structure from \cite{Rashidi2017_1-2-2-1} designed to illustrate an observed hopping behavior has also been reproduced in SiQAD as seen in \fref{\ref{fig:qsi_1-2-2-1}}. This structure has a degenerate ground state, with either of the two inner Si-DBs being occupied. It is observed that over a series of AFM line scans either of the inner Si-DBs tends to be fully occupied, and that state tends to persist over multiple line scans. In \cite{Rashidi2017_1-2-2-1}, this behavior is attributed to stabilization of the charge states by lattice relaxation \cite{Schofield2013,Kawai2016}, presenting an additional energetic barrier between the degenerate states. The tip influence and thermal fluctuations likely contribute to overcoming this barrier.

The potential to construct nanometer-scale logic circuits from atomic Si-DBs makes this technology a promising candidate for advancing the work in circuit miniaturization and power reduction. Further, the use of silicon substrate opens up possibilities to integrate Si-DB circuits with the existing silicon CMOS-based infrastructures. This is a new design space with largely unknown design rules and constraints; many more logic gates and circuit components still await investigation. The SiQAD project aims to reduce the barrier for the research community to explore this design space.

\section{SiQAD Overview} \label{sec:siqad_overview}

SiQAD is a CAD tool that enables the design and simulation of Si-DB circuits through an intuitive graphical user interface (GUI) and a modular simulation back-end. The GUI is written in C++ with the Qt cross-platform application framework \cite{Qt} and provides tools for Si-DB layout design, simulation management, as well as the placement of electrodes. Simulation engines are built on a standard framework with support for either C++ or Python.

Three simulation tools are included with the initial release of SiQAD: SimAnneal, a simulated annealer for finding low energy charge configurations; HoppingDynamics, a non-equilibrium electron hopping dynamics simulator intended for capturing the experimentally observed hopping behavior; and PoisSolver, an electrostatics solver which solves the generalized Poisson equation to find the electric potential landscape arising from any electrodes. All simulation engines exchange information with the SiQAD GUI using a standardized application programming interface (API). Additional engines can be implemented and incorporated through the API. 

The source code of the GUI and simulation tools is publicly available. The open-sourced tools allow the research community to modify or contribute additional functionalities, supporting the future growth of SiQAD as an Si-DB circuit design platform.

\section{Simulation Engines} \label{sec:simulation_engines}

\newcommand{\kt}{k_BT}

Three simulation engines are currently available with SiQAD, each with different purposes and capabilities.

\subsection{SimAnneal: Low Energy Charge Configuration Simulator}

\subsubsection{Working Principle}
\begin{figure}
  \centering
  \includegraphics[width=\linewidth]{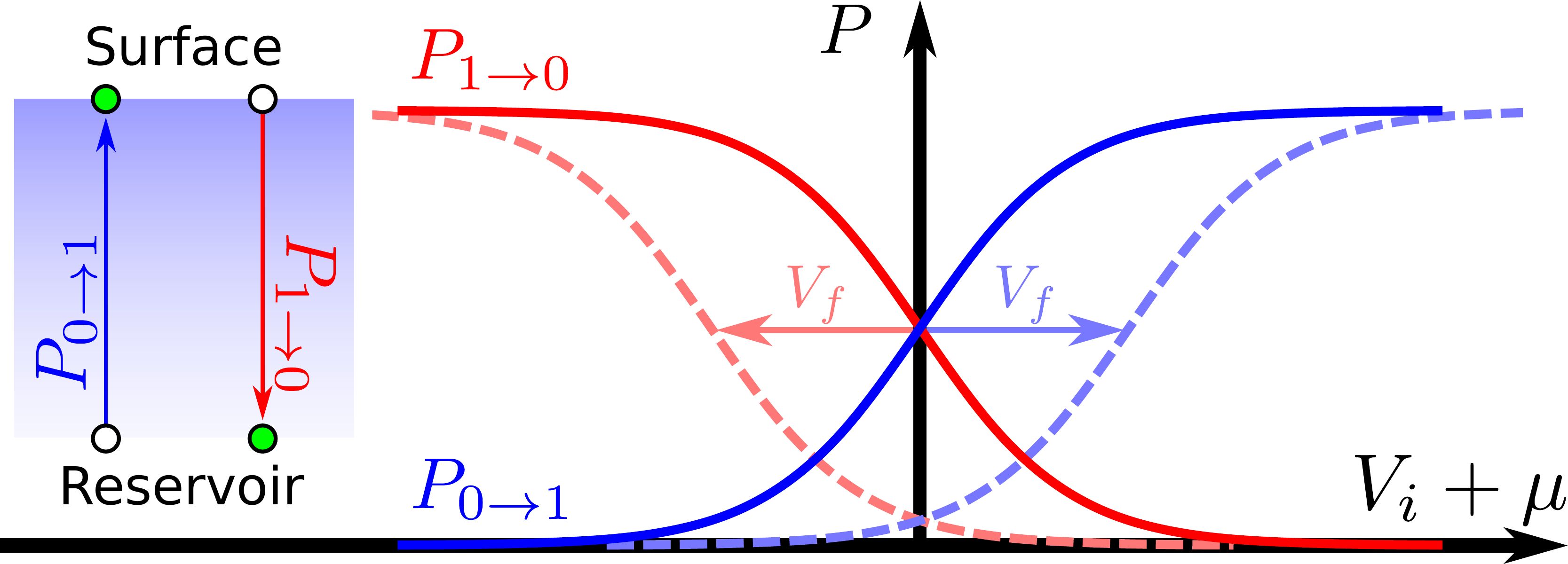}
  \caption{Fermi-Dirac distributions for electrons to tunnel on ($P_{0 \rightarrow 1}$) or off ($P_{1 \rightarrow 0}$) Si-DB $i$ as a function of the local potential. Charges hop between the Si-DBs on the surface and an imaginary reservoir that can provide or contain any number of electrons. As the freeze-out potential, $V_{f}$, is increased, it becomes harder for charges to hop between the Si-DBs and the reservoir.}
  \label{fig:simanneal_hop_sigmoid}
\end{figure}

\newcommand{\vext}{V^{\text{ext}}}
\newcommand{\vfz}{V_f}
SimAnneal is a simulated annealing algorithm containing two major components: \textit{population control} and \textit{surface hopping}.  The energy of a charge configuration is given by
\begin{equation}
 E(\vec{n}) = - \sum_i \vext_i n_i + \sum_{i<j} V_{ij} n_i n_j,
\end{equation}
where $\vext_i$ is the total contribution of external potential sources with the exception of influences from other Si-DBs at the i\tts{th} Si-DB, $n_i+1$ is the charge occupation of each Si-DB, and $V_{ij}$ is the strength of the Coulombic interaction between each Si-DB pair. During the surface hopping phase, attempts are made to hop charges from DB- occupied sites to DB0 sites with these hops accepted according to an acceptance function. A screened Coulomb potential is used, of the form
\begin{equation}
 V_{ij} = \frac{q_0}{4\pi\epsilon} \frac{1}{r_{ij}} \erf\left(\frac{r_{ij}}{\sigma_G}\right) e^{-\sfrac{r_{ij}}{}\lambda_{TF}},
\end{equation}
with $r_{ij}$ the distance between the Si-DBs, $\sigma_G$ the width of a Gaussian cloud approximation of the occupying charge, and $\lambda_{TF}$ the Thomas-Fermi screening length. In the results presented in this paper, we take $\sigma_G = \SI{0.5}{\nm}$, $\lambda_{TF} = \SI{5}{\nm}$, and $\epsilon_r = 6.35$. Each Si-DB experiences an effective local potential, $V_i$, given by
\begin{equation} \label{eq:simanneal_v_local}
 V_i = \vext_i - \sum_j V_{ij}n_j,
\end{equation}
which is assumed to further contribute to band bending. For each iteration of SimAnneal, we compute these local potentials at each site. As illustrated in \fref{\ref{fig:simanneal_hop_sigmoid}}, charges are removed from DB- sites with probability $P_{1 \to 0}$ and added to DB0 sites with probability $P_{0 \to 1}$:
\begin{subequations}
\label{eq:sa-pop}
\begin{align}
 P_{0 \to 1}(V_i) &= \frac{1}{1+e^{\sfrac{-(V_i+\mu-\vfz)}{\kt}}},\\
 P_{1 \to 0}(V_i) &= \frac{1}{1+e^{\sfrac{(V_i+\mu+\vfz)}{\kt}}}
\end{align}
\end{subequations}
Here the chemical potential $\mu$ is an empirical measurement of the Fermi level with respect to the \mbox{DB(0/-)} transition level of an isolated Si-DB, $\vfz$ is the \textit{freeze out} potential which is increased each iteration and tends to fix the number of electrons in the surface, and $k_B T$ is an artificial thermal energy that decreases for each iteration. The empirical chemical potential is intended to capture any band bending effects that affect all Si-DBs approximately equally, such as those due to inhomogeneity in the dopant concentration, to allow the tool to account for certain variation in device fabrication.

\subsubsection{Simulation Results}

The OR gate from \cite{Huff2017_OR} and the six Si-DB symmetric structure from \cite{Rashidi2017_1-2-2-1} have been reproduced and simulated in SiQAD.  \fref{\ref{sfig:simanneal_or}} shows the simulated ground states of the OR gate with various inputs. To acquire OR gate behavior, a minor modification had to be made from the design in \cite{Huff2017_OR}: the output perturber had to be moved one dot lower to weaken its influence. The simulated ground states all exhibit OR gate behavior with the correct outputs corresponding to the inputs. The output perturber modification and minor differences in electron population may be attributed to the absence of an AFM tip in the simulated results.

\fref{\ref{fig:qsi_1-2-2-1}} shows the simulated ground states of the six Si-DB symmetric structure. The perturbers on the two sides are always doubly-occupied, exerting an inward bias on the electron occupying the Si-DBs in between. One electron is equally shared between the two degenerate inner Si-DBs. This is in agreement with the experimental results of \cite{Rashidi2017_1-2-2-1}, where the middle electron hops between those two Si-DBs.

\subsection{HoppingDynamics: Non-equilibrium Dynamics Simulator}

\subsubsection{Working Principle}

\newcommand{\dg}[1]{\Delta G_{#1}}
\newcommand{\nug}{\eta}

HoppingDynamics is a non-equilibrium electron dynamics simulator that treats electron transitions via hopping rates. The simulator is generalized to allow for various forms of hopping rates depending on the choice of hopping model. Both Mott Variable Range hopping \cite{Apsley1974} and Marcus hopping \cite{Marcus1993} have been implemented. The rate of hopping for a charge from the $i$\tts{th} (DB-) to $j$\tts{th} (DB0) Si-DB is taken to be of the form
\begin{equation} \label{eq:hopping-rates}
  \nu_{ij} \propto e^{-2\alpha r_{ij}} \nug(\dg{ij}),
\end{equation}
with $r_{ij}$ the distance between the Si-DBs, $\dg{ij}$ the change in Gibbs free energy associated with the hop, $\alpha$ the spatial decay of the hopping rate, and $\eta$ some function of $\dg{ij}$. If the average hopping frequency, $\nu_0$, between two degenerate Si-DBs at some distance $r_0$ is known, the hopping rates can be expressed as
\begin{subequations}
 \begin{align}
  \text{Mott}   &:&  \nu_{ij} &= \nu_0 e^{-2\alpha(r_{ij}-r_0)}e^{-\sfrac{\dg{ij}}{\kt}}\\
  \text{Marcus} &:&  \nu_{ij} &= \nu_0 e^{-2\alpha(r_{ij}-r_0)}e^{-\sfrac{\dg{ij}(\dg{ij}+2\lambda)}{4\lambda\kt}}
 \end{align}
\end{subequations}
with $\lambda$ a measure of the energy reduction in occupied DB- states associated with the stabilization from the lattice. At each instant in time, a hop from $i$ to $j$ occurs with probability $P_{i\to j} = \nu_{ij}dt$. A spatial decay of $\alpha = \SI{1}{\per\nm}$ was assumed, and $\lambda$ calibrated to the experimental results with $\nu_0$ and $r_0$ taken from the combined AFM line scans in \cite{Rashidi2017_1-2-2-1}. Charge population can be modelled by adding additional channels for hopping to and from the surface Si-DBs. As an example, charges hop from DB- sites to the bulk and from the bulk to DB0 sites with respective rates $\nu_{i,B}$ and $\nu_{B,i}$:
\begin{subequations}
 \begin{align}
  \nu_{i,B} &= \nu_B f(V_i+\mu+\lambda),\\
  \nu_{B,i} &= \nu_B f\left(-(V_i+\mu)\right)
 \end{align}
\end{subequations}
Where $\nu_B$ is some characteristic rate and $f$ some function of the energy difference between the Fermi level and the \mbox{DB(0/-)} transition level. In these results, $f$ is taken to be a logistic function and $\nu_B$ sufficiently large that surface-bulk transitions are effectively immediate once DB- states pass the Fermi level.

\subsubsection{Simulation Results}

The six Si-DB symmetric structure from \cite{Rashidi2017_1-2-2-1} has been simulated with HoppingDynamics. Si-DB charge states were measured according to a tip scanning at $\SI{8.9}{\nm\per\second}$ for $800$ line scans to match the experimental time scales with the tip influence ignored. An example result is shown in \fref{\ref{fig:qsi_1-2-2-1}}. Like the experimental results from \cite{Rashidi2017_1-2-2-1}, the structure always contains 3 electrons with similar hopping between the two degenerate Si-DBs. The OR gate results were also simulated. Identical results to SimAnneal were achieved for the first three input cases. For the $11$ input, similar hopping between the two half filled Si-DBs in \fref{\ref{sfig:simanneal_or}} was observed. These results are not shown to avoid repetition.

\subsection{PoisSolver: Electrostatics Solver}

\subsubsection{Working Principle}
Electron occupation on the surface may be controlled by shifting the DB transition levels with respect to the bulk Fermi level. Currently, the Fermi levels can be controlled either by doping, which cannot be changed after fabrication, or AFM tip-induced band bending, which requires delicate equipment setup. We envision that future Si-DB circuits will rely on buried or suspended electrodes with controllable potentials to adjust the surface electron population, allowing fine control of information flow. Sections of the circuit may be switched `off' by applying a voltage that induces upward band bending on the surface until the contained Si-DBs reach the DB0 state; they may be switched `on' by inducing downward band bending until the desired electron population has been reached. In order to better understand the behavior of the Si-DBs under the influence of electrodes, the effect of the electrodes on the surface potential must be quantified.

An FEM solver for the generalized Poisson equation was implemented in Python using the FEniCS software package \cite{AlnaesBlechta2015a, LoggMardalEtAl2012a,LoggWells2010a,LoggWellsEtAl2012a,KirbyLogg2006a,LoggOlgaardEtAl2012a,OlgaardWells2010b}. The engine is able to account for user-defined design parameters, most notably the geometrical dimensions of the sample and the electrodes, the resolution of the finite element mesh, and the dielectric permittivity of the simulated region.

Once the parameters are defined and the simulator is invoked, a mesh is defined and generated according to the provided geometry and resolution, and the generalized Poisson equation is solved over the mesh. Variable mesh generation is implemented such that the mesh is generated with finer detail at material boundaries (e.g.\ electrodes and silicon surface), while other areas will have a coarser mesh. The mesh definition is then passed to Gmsh to generate the appropriately formatted mesh for simulation \cite{Geuzaine2009}. When the computations are complete, the potentials are written to a file which can be used in other engines. A two dimensional slice of the result is created and displayed to the user in a color map. For time-varying clocking fields, the electrode potentials are adjusted with each simulation time step and PoisSolver is used to recompute the potential landscape.

\subsubsection{Simulation Results}
\begin{figure}
  \centering
  \includegraphics[width=\linewidth]{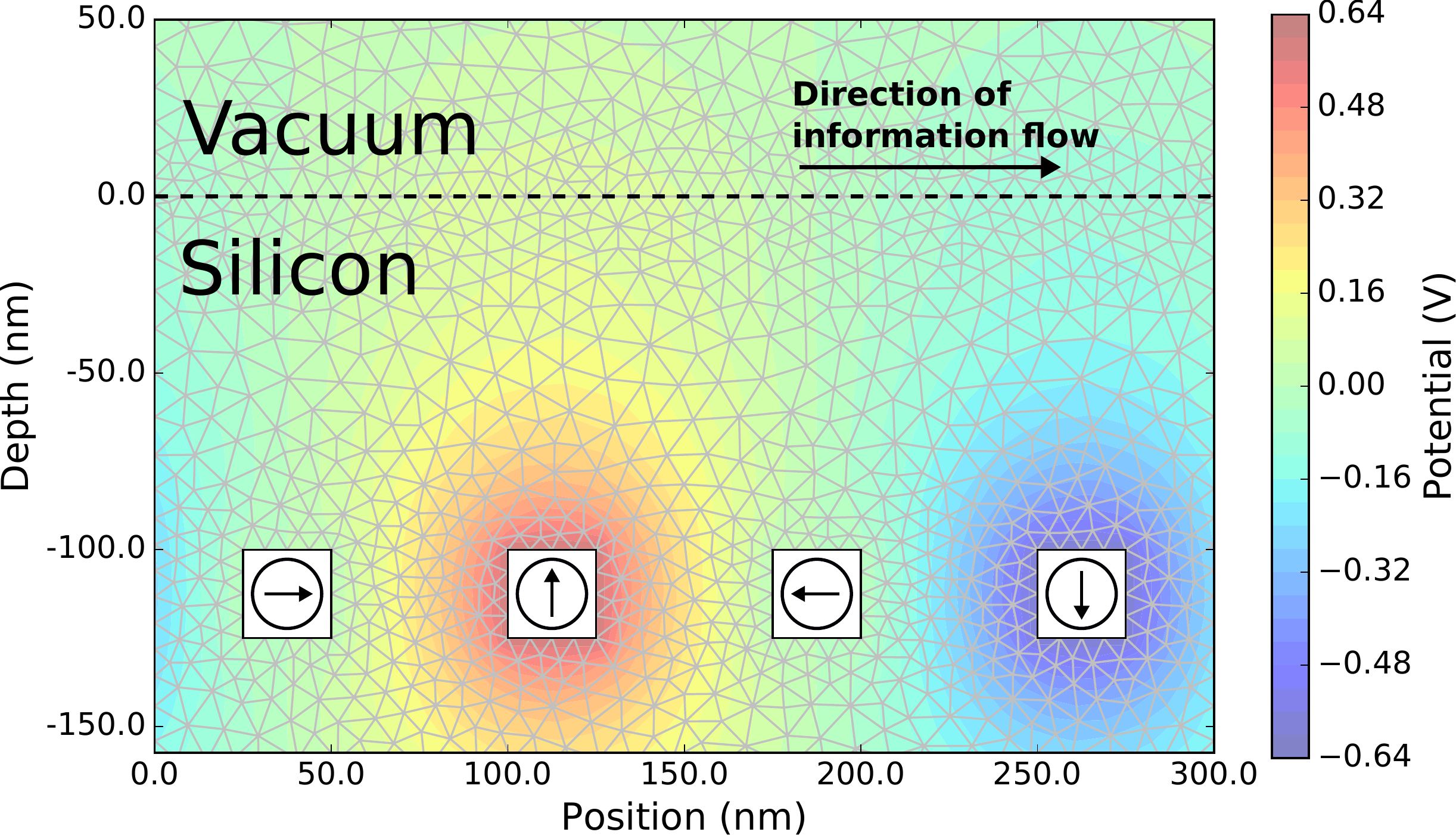}
  \caption{A schematic of a buried electrode system with dimensions that are realistic for fabrication. The mesh was generated by PoisSolver, which was then used to solve for the potential landscape of the cross-section. The electrodes (white squares) have time-varying potentials in the form of sinusoidal functions, each with a $90^\circ$ phase offset from the nearest one to the left. These phases are indicated by the phasor inside each electrode.}
  \label{fig:poissolver_buried_electrodes}
\end{figure}

\begin{figure}
  \centering
  \includegraphics[width=\linewidth]{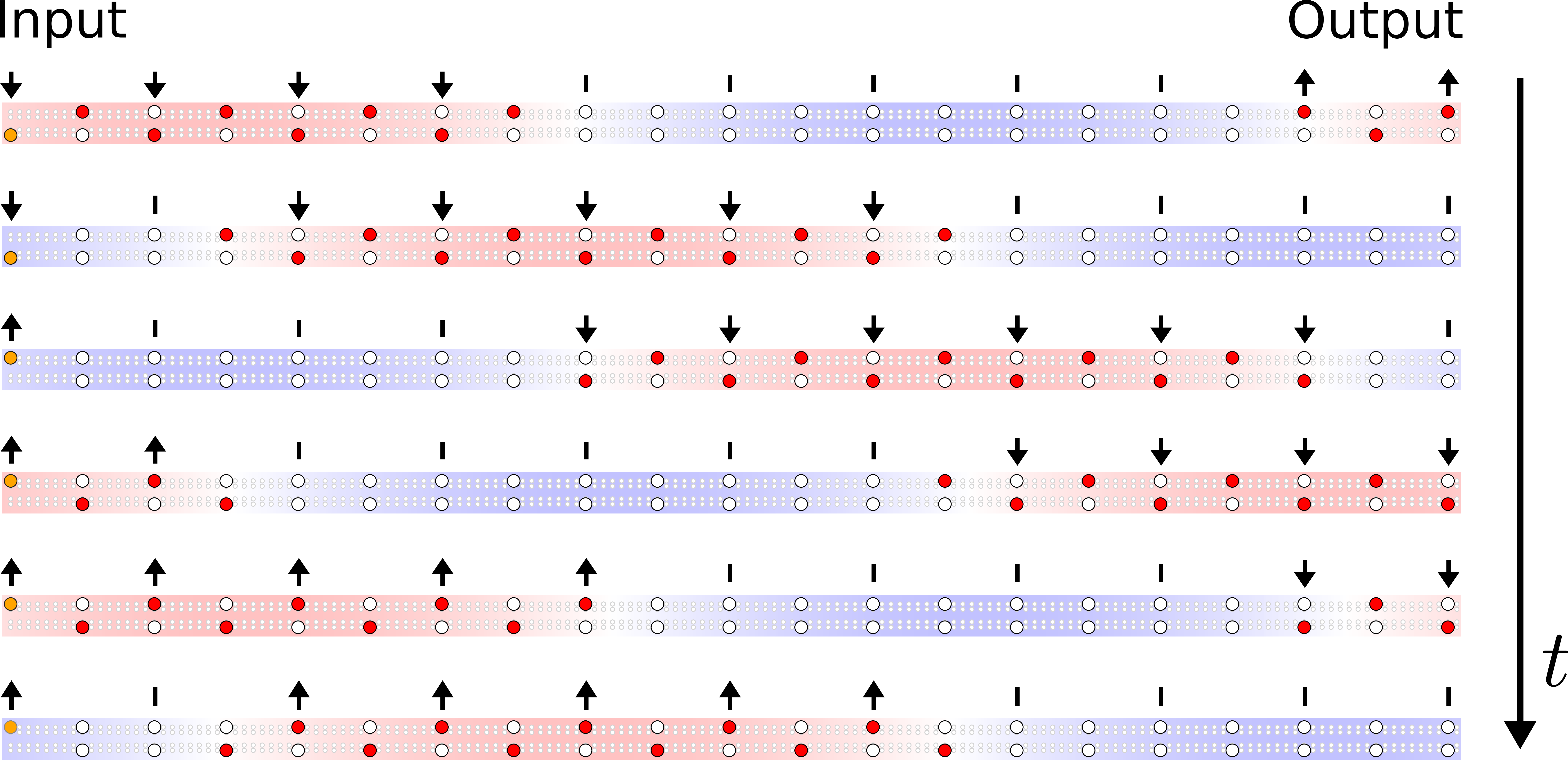}
  \caption{Signal packet propagation using population clocking in an inverting wire. Shading indicates the $\pm \SI{100}{\meV}$ surface potential waveform achieved using 4-phase electrodes $\SI{100}{\nm}$ below the surface. The arrows indicate the logical state of the inverting wire. Note that by changing the input perturber during the depopulated phase two signal packets of opposite polarization are generated. To visualize the dots and packets for illustration, it was necessary to use electrodes $\SI{4}{\nm}$ wide and separated by $\SI{8}{\nm}$. Similar results are achieved for larger electrodes.}
  \label{fig:clocked_wire}
\end{figure}

In order to investigate clocked information flow via buried electrodes, an inverting Si-DB wire structure was implemented in SiQAD with sinusoidally clocked 4-phase electrodes located $\SI{100}{\nm}$ below the surface, as illustrated in \fref{\ref{fig:poissolver_buried_electrodes}}. The mesh generated by PoisSolver and the electric potential of the cross-section are shown. An electrode clocking amplitude of $\SI{0.6}{V}$ was chosen to produce a $\pm\SI{100}{\meV}$ surface potential waveform to provide the right amount of band bending to control Si-DB states between DB0 and DB-. Simulation results of the clocked inverting wire using HoppingDynamics can be seen in \fref{\ref{fig:clocked_wire}}.

\section{Simulating New Logic Designs} \label{sec:new_gates}

\begin{figure}
  \centering
  \subfloat[AND gate.]{
    \includegraphics[width=\linewidth]{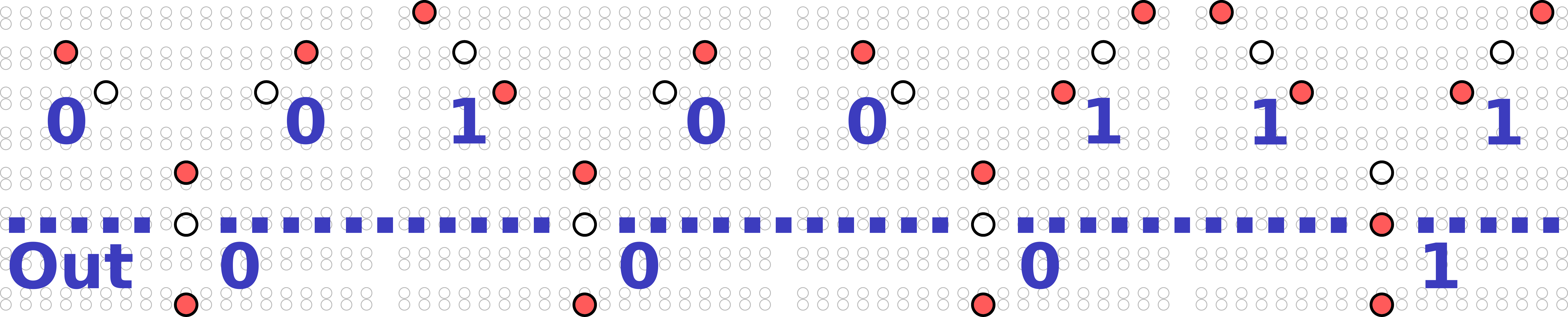}
    \label{fig:simanneal_and_gate}}\hfill
  \subfloat[XOR gate.]{
    \includegraphics[width=\linewidth]{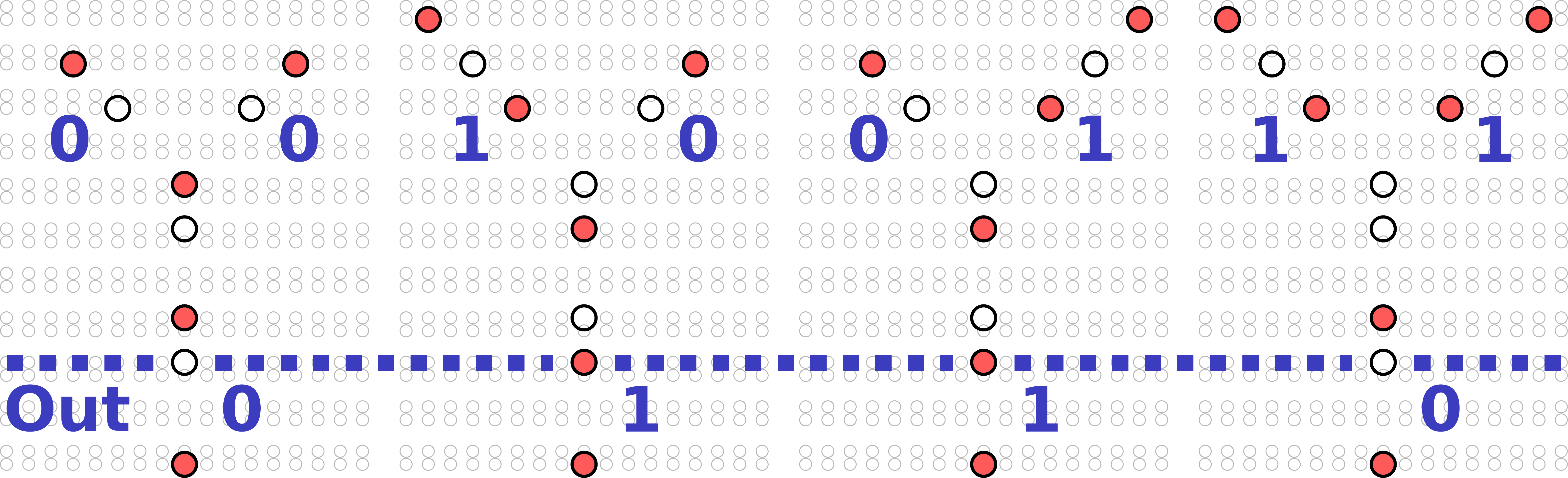}
    \label{fig:simanneal_xor_gate}}\hfill
  \subfloat[NAND gate.]{
    \includegraphics[width=\linewidth]{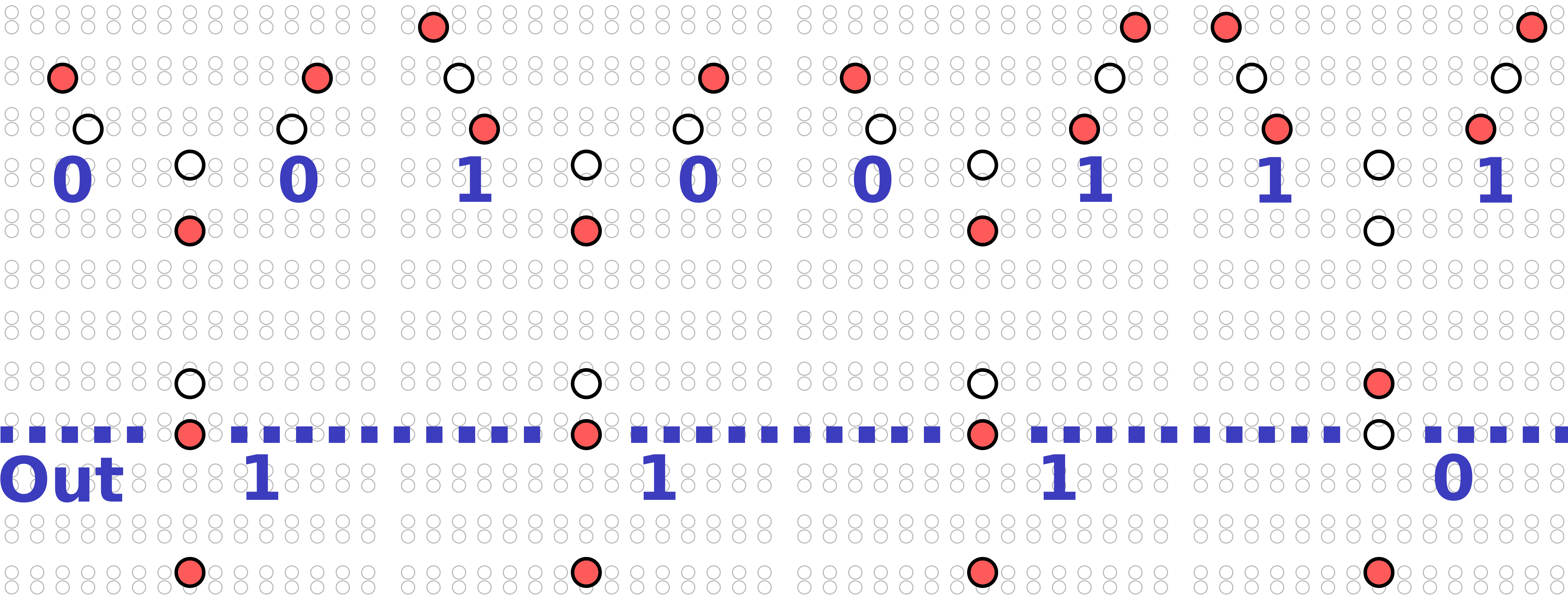}
    \label{fig:simanneal_nand_gate}}\hfill
  \subfloat[NOR gate.]{
    \includegraphics[width=\linewidth]{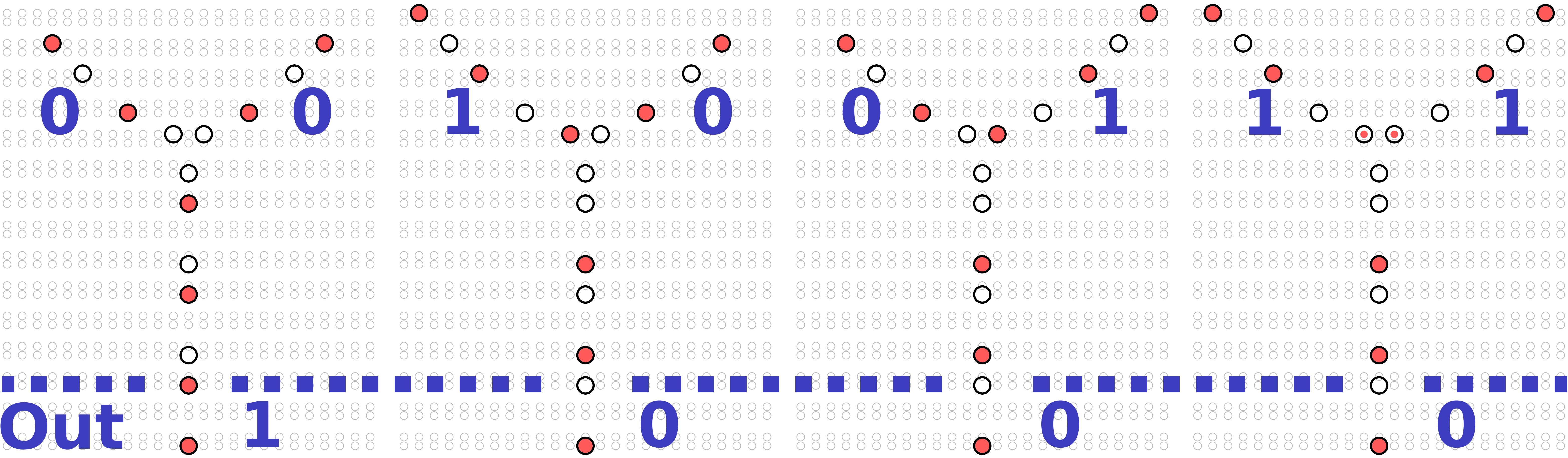}
    \label{fig:simanneal_nor_gate}}\hfill
  \caption{New logic gates designed in SiQAD and simulated in SimAnneal with $\mu=\SI{0.25}{\eV}$. The inputs are at logic `0' by default; the addition of perturbers force those inputs to logic `1'. All possible input configurations have been shown. %
    \bpar{a} 2-input AND gate. %
    \bpar{b} 2-input XOR gate. %
    \bpar{c} 2-input NAND gate. %
    \bpar{d} 2-input NOR gate. %
  }
\end{figure}

Using SiQAD, new Si-DB logic gates have been designed and simulated. Like the OR gate, these logic gates represent bit information through the locations of charges in pairs of Si-DBs. The inputs are at logic `0' by default, and are pushed to logic `1' in the presence of perturbers. Simulation results of the ground state configurations for the 2-input logic AND, XOR, NAND and NOR gates are illustrated in \fref{\ref{fig:simanneal_and_gate}}-\ref{fig:simanneal_nor_gate} respectively with correct logic functionalities.

\section{Conclusion} \label{sec:conclusion}

Motivated by recent developments in Si-DB quantum dots which have shown their potential to serve as emerging building blocks for nano-scale logic circuits \cite{Wolkow2014_logic, Huff2017_OR, Rashidi2017_1-2-2-1}, SiQAD has been developed to enable the design and simulation of Si-DB circuits. Three included simulation engines have been introduced: SimAnneal, a ground state electron configuration finder; HoppingDynamics, a non-equilibrium electron dynamics simulator; and PoisSolver, an electrostatics solver which finds the global electric potential landscape by solving the generalized Poisson equation. An OR gate proposed and investigated in \cite{Huff2017_OR} and a six Si-DB symmetric structure from \cite{Rashidi2017_1-2-2-1} have been recreated in SiQAD and simulated with SimAnneal to show similar ground-state results. The six Si-DB symmetric structure has also been simulated with HoppingDynamics to show comparable electron hopping behavior amongst degenerate states.

Using SiQAD's design capabilities, an AND gate, XOR gate, NAND gate and NOR gate constructed of Si-DBs have been proposed and simulated, laying the groundwork for future Si-DB logic designs. Simulation results for a clocked inverting wire have been presented, with PoisSolver solving the time-varying field generated by buried electrodes and HoppingDynamics simulating the system's dynamic behavior under the influence of that field. This demonstrates the tool's potential for investigating future electrode-clocked systems.

SiQAD's design and simulation functionalities will continue to be refined and improved. We encourage readers to follow this project's development at \cite{SiQAD}.

\section{Acknowledgments}

K.W.\ initiated and guided the development of SiQAD. S.N.,\ J.R.\ and H.N.C.\ developed SiQAD's GUI, simulation engines, and prepared the manuscript. R.L.\ contributed improvements to SimAnneal and designed logic circuits. M.R., W.V., T.D., L.L., R.W.\ and K.W.\ contributed towards the development of physics models. All authors reviewed and commented on the manuscript. We would like to thank NSERC for their financial support.

\bibliographystyle{ieeetr}
\bibliography{paper.bib}

\end{document}